\documentclass[12pt,times]{elsarticle}

\usepackage{amssymb}

\usepackage{lineno}

\usepackage{verbatim}

\newcounter{bla}

\journal{Computer Physics Communications}

\begin{document}

\begin{frontmatter}



\title{Parallel grid library for rapid and flexible simulation development}


\author[a,b]{Honkonen, I.\corref{author}}
\author[a]{von Alfthan, S.}
\author[a]{Sandroos, A.}
\author[a]{Janhunen, P.}
\author[a]{Palmroth, M.}

\cortext[author]{
Corresponding author.\\
\textit{E-mail address:} ilja.honkonen@fmi.fi\\
\textit{Tel:} +358503803147\\
\textit{Fax:} +358295394603\\
\textit{Postal address:} P.O. Box 503, 00101 Helsinki, Finland\\
}

\address[a]{Finnish Meteorological Institute, Helsinki, Finland}
\address[b]{Department of Physics, University of Helsinki, Helsinki, Finland}

\begin{abstract}

We present an easy to use and flexible grid library for developing highly scalable parallel simulations.
The distributed cartesian cell-refinable grid (dccrg) supports adaptive mesh refinement and allows an arbitrary C++ class to be used as cell data.
The amount of data in grid cells can vary both in space and time allowing dccrg to be used in very different types of simulations, for example in fluid and particle codes.
Dccrg transfers the data between neighboring cells on different processes transparently and asynchronously allowing one to overlap computation and communication.
This enables excellent scalability at least up to 32 k cores in magnetohydrodynamic tests depending on the problem and hardware.
In the version of dccrg presented here part of the mesh metadata is replicated between MPI processes reducing the scalability of adaptive mesh refinement (AMR) to between 200 and 600 processes.
Dccrg is free software that anyone can use, study and modify and is available at https://gitorious.org/dccrg.
Users are also kindly requested to cite this work when publishing results obtained with dccrg.

\end{abstract}


\end{frontmatter}

\linenumbers


{\bf PROGRAM SUMMARY}

\begin{small}
\noindent
{\em Manuscript Title:} Parallel grid library for rapid and flexible simulation development\\
{\em Authors:} Honkonen, I., von Alfthan, S., Sandroos, A., Janhunen, P., Palmroth, M.\\
{\em Program Title:} DCCRG                                    \\
{\em Journal Reference:}                                      \\
{\em Catalogue identifier:}                                   \\
{\em Licensing provisions:} GNU LGPL v3                       \\
{\em Programming language:} C++                               \\
{\em Computer:} PC, Cluster, Supercomputer                    \\
{\em Operating system:} POSIX                                   \\
{\em RAM:} 10 MB - 10 GB per process                                              \\
{\em Number of processors used:} 1 - 32768 cores                             \\
{\em Supplementary material:}                                 
{\em Classification:} 4.12, 4.14, 6.5, 19.3, 19.10, 20             \\
{\em External routines/libraries:} MPI-2 [1], boost [2], Zoltan [3], sfc++ [4]        \\
{\em Nature of problem:} \\
  Grid library supporting arbitrary data in grid cells, parallel adaptive mesh refinement, transparent remote neighbor data updates and load balancing.
   \\
{\em Solution method:}\\
  The simulation grid is represented by an adjacency list (graph) with vertices stored into a hash table and edges into contiguous arrays.
  Message Passing Interface standard is used for parallelization.
  Cell data is given as a template parameter when instantiating the grid.
   \\
   \\
   \\
{\em Restrictions:} \\
  Logically cartesian grid.
   \\
{\em Additional comments:}\\
   \\
{\em Running time:}\\
  Running time depends on the hardware, problem and the solution method.
  Small problems can be solved in under a minute and very large problems can take weeks.
  The examples and tests provided with the package take less than about one minute using default options.
  In the version of dccrg presented here the speed of adaptive mesh refinement is at most of the order of $10^6$ total created cells per second.
   \\

\end{small}

\section{Introduction}
\label{sec:introduction}

During the rising phase of the solar cycle, it is becoming more important to understand the physics of the near-Earth space.
The dynamical phenomena caused by the constant flow of magnetized collisionless plasma from the Sun creates space weather that may have harmful effects on space-borne or ground-based technological systems or on humans in space.
While the physics of space weather is being studied with in situ instruments
(e.g. NASA's Radiation Belt Storm Probes launched in 2012-08-30\footnote{http://www.nasa.gov/mission\_pages/rbsp/main/index.html})
and by means of remote sensing, it is also important to model the near-Earth space with numerical simulations.
The simulations can be used both as a context to the one-dimensional data sets from obsevations, as well as a source to discover new physical mechanisms behind observed variations.
Present large scale (global) simulations are based on computationally light-weight simplified descriptions of plasma, such as magnetohydrodynamics (MHD, \cite{pekka12}, \cite{lyon04}, \cite{powell99} and \cite{raeder95}).
On the other hand the complexity and range of spatial scales (from less than $10^1$ to over $10^6$ km) in space weather physics signifies the need to incorporate particle kinetic effects in the modeled equation set in order to better model, for example, magnetic reconnection, wave-particle interactions, shock acceleration of particles, ring current, radiation belt dynamics and charge exchange (see e.g. \cite{hannu11} for an overview).
However, as one goes from MHD towards the full kinetic description of plasma (from hybrid PIC \cite{esa03} and Vlasov \cite{minna12} to full PIC \cite{birdsall}, \cite{hockney}), the computational demands increase rapidly, indicating that the latest high performance computing techniques need to be incorporated in the design of new simulation architectures.

As the number of cores in the fastest supercomputers increases exponentially the parallel performance of simulations on distributed memory machines is becoming crucial.
On the other hand, utilizing a large number of cores efficiently in parallel is challenging especially in simulations using run-time adaptive mesh refinement (AMR).
This is largely a data structure and an algorithm problem albeit specific to massively parallel physical simulations running on distributed memory machines.

In computer simulations dealing with, for example, continuous matter (a fluid) the simulated domain is discretized into a set of points or finite volumes which we will refer to as cells.
At any given cell the numerical solution of a differential equation describing the problem often depends only on data within a (small) part of the simulated volume.
This is true for a single time step in a solver for a hyperbolic problem or a single iteration in a solver for an elliptic problem.
This spatial data dependency can be implemented implicitly in the solver function(s) or explicitly as a separate grid library used by the application.

In a simple case the number of cells in the simulation stays constant and the data dependency of each cell is identical allowing cell data to be stored in an array whose size is determined at grid creation and the spatial neighbors to be represented as indices into this array.
A straightforward AMR extension of this concept is to create additional nested grids in specific parts of the simulation domain with higher resolution.
By solving each grid separately and interpolating the results from finer grids into coarser grids one does not have to modify the solver functions.
This technique is used extensively for example by Berger (see \cite{berger89} for some of the earliest work) and by \cite{henshaw08} and \cite{kohn01}.
In the rest of this work however we will concentrate on AMR implementations in which additional overlapping grids are not created but instead cells of the initial grid are refined, i.e. replaced with multiple smaller cells.

\begin{figure}[t]
\vspace*{2mm}
\begin{center}
\includegraphics[width=\columnwidth]{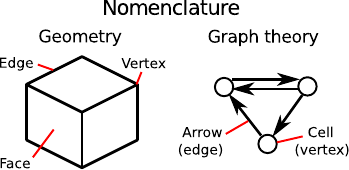}
\end{center}
\caption{
The nomenclature used in this work for geometry and graph theory.
Standard graph theoretical terms are given in parentheses for reference.
}
\label{fig:nomenclature}
\end{figure}

A generic unstructured grid (as provided for example by libMESH \cite{kirk06}) does not admit as simple a description as above and is generally described by a directed graph in which vertices represent simulation cells and directed edges represent the data dependencies between cells.
Unfortunately the nomenclature of graph theory and geometry overlap to some extent and discussing both topics simulataneously can lead to confusion.
Figure \ref{fig:nomenclature} shows the nomenclature we use from this point forward, the standard graph theoretical terms are given in parentheses for reference.
A cell is a natural unit in simulations using the finite volume method (FVM) and hereinafter we will use the term cell instead of vertex when discussing graphs.
Also an edge in FVM simulations usually refers to the edges of a cube representing the physical volume of a cell, and hence we will use the term arrow to refer to a directed edge in a graph.
Furthermore we note that each cell in the grid can also represent, for example, a block of cells similarly to \cite{powell99}, but for the purposes of this work the actual data stored in grid cells is largely irrelevant.

Since a graph can also be used to represent the cells and arrows of grids simpler than an unstructured mesh, the question arises how does a particular program implement its graph representation of the simulated system, e.g.~what simplifying assumptions have been made and how is the graph represented in memory.
A popular representation in (M)HD AMR simulations is to have a fixed number of arrows directed away from each source cell and to store the arrows as native pointers to the destination cells.
In case a cell does not exist all arrows pointing to it are invalidated in neighboring cells.
This technique has been used with different variations by \cite{holst07}, \cite{khokhlov98}, \cite{macneice00} and \cite{stout97}, for example.

There are several possibilities for representing the cells and arrows of a graph, for example an adjacency list or an adjacency matrix \cite{cormen01}.
In physical simulations the number of arrows in the graph is usually of the same order as the number of cells in which case a suitable representation is an adjacency list.
In an adjacency list the cells of the graph are separate objects and each cell stores the arrows pointing to and/or from that cell.
The cells of the graph and the arrows of each cell can be stored in different types of data structures.
For example the cells are stored in a contiguous array (representing a linear octree) in \cite{hariharan05}, \cite{sundar08}, \cite{burstedde11} and \cite{bangerth11}, a hash table in \cite{warren93} and a (doubly) linked list in \cite{holst07}.
On the other hand the arrows of each cell are stored in a fixed size array of native pointers in \cite{holst07} and as single bits in \cite{warren93}.

In this work we introduce the distributed cartesian cell-refinable grid (dccrg) for rapid development of parallel simulations using, for example, finite volume or finite element methods (FEM).
In dccrg the graph is represented by an adjacency list in which cells are stored into a hash table, while the arrow lists of cells are stored into contiguous arrays.
We describe the details of the graph representation in Section \ref{sec:hash_table}.
In section \ref{sec:implementation} we describe the C++ implementation of dccrg and present its unique features with respect to other published grid codes:
arbitrary data in grid cells, transparent updates of remote neighbor data, user-selectable neighborhood size for cells and ease of use.
In section \ref{sec:tests} we test the scalability of dccrg using a variety tests in one, two and three dimensions and draw our conclusions in Section \ref{sec:discussion}.

\section{Implementation of the grid graph}
\label{sec:hash_table}

Dccrg represents the grid as an adjacency list in which cells are stored into a hash table.
A hash table has one clear advantage over a tree when used to store the grid:
cells can be accessed, inserted and deleted in constant amortized time regardless of the number of cells and their physical size and location.
Thus neither the total number of cells nor the number of refinement levels affect the simulating performance of a single core.
Each cell is associated with a unique id which we use as a key into the hash table.
A potential drawback of a hash table is the computational cost of the hash function, but according to our tests the cost is usually not important.
The time to solve one flux between two cells in the MHD tests presented in Section \ref{sec:mhd_tests} is about four times larger than accessing one random cell in the hash table.
The cell access time can be optimized further, for example, by storing and solving blocks of cells instead of single cells as is done in \cite{powell99} and discussed further in Section \ref{sec:arbitrary_data}.

\subsection{Mapping cell ids to a physical location}
\label{sec:cell_mapping}

Our cell ids are globally unique integers which offers several advantages:
1) The cell id can be calculated locally, i.e. without communication with other processes,
2) The neighbors of a cell can be stored as cell ids instead of pointers that are not consistent across computing nodes,
3) The cost of computing the hash function values is minimized,
4) The memory required for storing the cell ids is small.

\begin{figure}[t]
\vspace*{2mm}
\begin{center}
\includegraphics[width=\columnwidth]{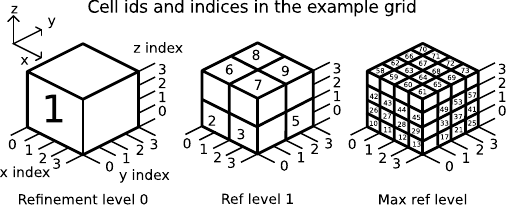}
\end{center}
\caption{
An example dccrg grid of size 1 in each dimension in cells of refinement level 0, with a maximum refinement level 2, showing the ids and indices of all possible cells.
}
\label{fig:example_grid}
\end{figure}

Figure \ref{fig:example_grid} shows an example of mapping cells to unique ids which is done as follows:
cell ids within each refinement level increase monotonically first in x coordinate, then in y and then in z, with cells at refinement level 0 represented by numbers from 1 to $N_0$, refinement level 1 by numbers from $N_0 + 1$ to $N_0 + 1 + N_1$, etc.
Cells at refinement level $l + 1$ are half the size of cells at refinement level $l$ in each dimension.
Cells at equal refinement level are identical in size.
Hence in three dimensions $N_0 = \frac{N_1}{8} = \frac{N_2}{64} = ... = n_x n_y n_z$, where $n_x, n_y$ and $n_z$ are the grid size in cells of refinement level 0 in the x, y and z dimensions respectively.
Hereinafter cell size refers to the logical size of cells assuming a homogeneous and isotropic cartesian geometry.

When searching for the neighbors of a cell in the hash table (see Section \ref{sec:neigh_search}) it is convenient to use the concept of cell indices:
the location of each cell in the grid is represented by one number per dimension in the interval $[0, 2^L n_i - 1]$ where L is the maximum refinement level of the grid and $n_i$ is $n_x$, $n_y$ or $n_z$ respectively.
Figure \ref{fig:example_grid} shows the possible cell indices for an example grid with $n_x = n_y = n_z = 1$ and $L = 2$.
The size of a single cell of refinement level $l$ is $2^{L - l}$ indices in each dimension.
A cell spanning more than one index is considered to be located at indices closest to the origin of the grid, for example cell \#3 in Figure \ref{fig:example_grid} is located at indices (2, 0, 0).
Similarly to \cite{keppens12} there is a one-to-one mapping between cell ids and cell indices plus refinement levels, e.g.~in addition to its id a cell can be uniquely identified by its indices and refinement level.

In the current implementation of dccrg a cell is refined by creating all of its children; in the example grid of Figure \ref{fig:example_grid} refining cell \#1 would create cells \#2...\#9.
In principle this is not required and more complex grid structures are possible in which, for example, the grid in Figure \ref{fig:example_grid} would consist of cells \#1 and \#3 alone.
Such an approach has been found useful by \cite{holst07}.
Complete refinement of cells in our case was a practical decision based on our current simulation needs and it also simplifies the neighbor searching code and enables optimizations described in the next section.

\subsection{Neighbor searching}
\label{sec:neigh_search}

\begin{figure}[t]
\vspace*{2mm}
\begin{center}
\includegraphics[width=\columnwidth]{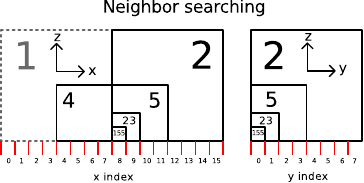}
\end{center}
\caption{
An example grid illustrating neighbor searching for cell \#4 in the positive x direction.
The siblings of cell \#4 are not shown for clarity and also some of its potential neighbors in positive x direction have been left out.
}
\label{fig:neighbors}
\end{figure}

In dccrg all cells existing within a certain minimum distance from local cells (cells owned by the process) are stored in a hash table with the cell id as the key and the process owning the cell as the value.
Since the mapping of cell ids to a location is unique, finding the neighbors of a cell in the hash table is straightforward:
for all indices neighboring a given cell the hash table is searched for cells of all applicable refinement levels.
Figure \ref{fig:neighbors} shows an example of neighbor searching in a grid with $n_x = 2, n_y = n_z = 1, L = 3$ and a neighborhood size of one.
The siblings of cell \#4 (\#3, \#7, \#8, \#11, \#12, \#15 and \#16) are not shown for clarity and some potential neighbors of cell \#4 in the positive x direction have also been omitted.
In dccrg the cells' neighborhoods are measured in multiples of their own size, e.g. for cell \#4 all cells that are (at least partially) between indices 0 and 11 inclusive in the x direction would be considered as neighbors.
In order to find the neighbor(s) of cell \#4 in positive x direction the hash table is searched for the smallest cell at indices (8, 0, 0) which in this case could correspond to any of the following cells: \#2, \#5, \#23 or \#155.
If cell \#23 is the smallest cell found in the hash table the search can stop since it is known that the siblings of cell \#23 also exist (not shown, cells \#24, \#31, \#32, \#55, \#56, \#63 and \#64) because all children of a cell are created when a cell is refined.
On the other hand if cell \#155 is the smallest cell found at indices (8, 0, 0) then the search would continue at indices outside of cell \#155 and its siblings, for example at indices (10, 0, 0) or (8, 2, 0).

The concept of hanging nodes or faces used in unstructured mesh codes is not directly applicable to dccrg because a single cell is the smallest unit that dccrg deals with.
Since the user is responsible for defining what is stored in each cell he/she must also define, if required, the data stored at the faces, edges and vertices of cells.
Hanging nodes, which are the result of cells of different size sharing an edge or a face, must be handled by the solvers used for a particular application.
For example in GUMICS-4 \cite{pekka12} where the magnetic field is separated into background and perturbed components the face average background magnetic field is required when solving the flux through a face.
With AMR a face average value of the background field is required for every face of every cell because, for example, if only one face average is stored per dimension in cells it would not be possible to solve the flux between cells (\#4, \#2) and (\#1, \#5) in Figure \ref{fig:neighbors}.
This is due to the fact that the face average value of the smaller cell must be used in both cases and it is only available if every face of every cell stores the face average field.

Currently dccrg enforces a maximum refinement level difference of 1 between neighboring cells.
Hence it is sufficient to search for cells of three refinement levels $l-1...l+1$ when finding the neighbors of a cell of refinement level $l$.
In principle the enforcement of maximum refinement level difference is not required.
For example in Figure \ref{fig:neighbors} cell \#4 (refinement level 1) has cell \#155 (refinement level 3) as a neighbor but in the current version of dccrg such a situation is not permitted and searching the hash table for cells \#2, \#5 and \#23 is sufficient for finding the neighbors of cell \#4.
This was a practical decision based on our experience with global MHD modeling of the Earth's magnetosphere using GUMICS-4.
In future this restriction might be removed.
A similar one is also used in \cite{keppens12}.

Even though a maximum refinement level difference of one is enforced between neighboring cells and searching for cells in the hash table is a quick operation, in practice the ids of neighbors of local cells are also stored explicitly.
As mentioned in Section \ref{sec:introduction}, this neighbor information corresponds to arrows between cells in a graph and hence we will use the term arrow list to refer to the neighbor list of a cell.
Dccrg stores both the arrows pointing away from and the arrows pointing to local cells.
With AMR it is possible that there exists only one arrow between two cells because cells' neighborhoods are measured in units of their own size.
For example in Figure \ref{fig:example_grid} with a neighborhood size of 1, cell \#13 would be considered a neighbor of cell \#2 but cell \#2 would not be considered a neighbor of cell \#13.
Explicitly storing the arrows to and from local cells enables fast iteration for example by user code (solvers, reconstruction functions, etc).
In dccrg the arrow lists of local cells are stored as contiguous arrays.

\subsection{Algorithmic advantages}
\label{sec:algo_adv}

The most important advantage that globally unique cell ids have over a traditional graph implementation using native pointers between cells is that the arrows between cells are not required to be consistent during AMR or load balancing.
For example when doing AMR, a pointer-based implementation has to be careful not to leave any dangling pointers and to update the pointers of all nearby cells in the correct order so as not to lose access to any cell.
On the other hand with unique cell ids the arrow lists of cells can simply be emptied when needed and new neighbors searched in the hash table as described in Section \ref{sec:neigh_search}.
In addition to being easy to implement this method also admits simple thread-based parallelization inside dccrg due to different threads modifying only the arrow lists of different cells.

It is also advantageous to use unique cell ids in arrow lists instead of pointers in a parallel program running on distributed memory hardware.
In this environment a pointer to a neighboring cell is only valid on one process and cannot be used to refer to the same neighbor on other processes.
On the other hand the same unique cell id can be used by all processes to refer to the same neighbor regardless of its actual location in memory.

\section{Implementation}
\label{sec:implementation}

A separate grid library is a natural abstraction probably for any physical simulation but especially for simulations using FVM where the concept of a grid and its cells' data dependencies are easy to define and implement.
Thus following good software development practice dccrg is implemented independently of any specific physical problem or its solver, while still providing the flexibility required for various types of simulations, for example (M)HD, advection (e.g.~Vlasov) and kinetic.

Dccrg is written in C++ which allows us to easily separate low level functionality of dccrg into subclasses which higher-level classes can use thus also benefiting from a modular internal implementation, a technique also used in \cite{kirk06}.
For example the physical geometry of the grid is handled by a separate class which is also given to dccrg as a template parameter.
This allows one to easily extend the grid geometries supported by dccrg.
In the default homogeneous and cartesian geometry cells of the same refinement level are identical in size in each dimension\footnote{https://gitorious.org/dccrg/dccrg/blobs/master/dccrg\_constant\_geometry.hpp}.

Here we describe the unique user-visible features of dccrg with respect to other grid codes and also present important features of the serial and parallel implementation.

\subsection{Unique features}

\subsubsection{Arbitrary data in grid cells}
\label{sec:arbitrary_data}

The most important feature distinguishing dccrg from other grid codes is the possibility of trivially storing data of arbitrary type and size in the grid's cells by simply giving the class which is stored in the cells as a template parameter to dccrg when creating an instance of the grid.
This also allows a single simulation to have several independent parallel grids with different geometries and different types of data stored in the grids' cells'.
The amount of data can also vary between different cells of the same grid and in the same cell as a function of time.
This is required for example in kinetic simulations where not only does the total number of particles change but also the number of particles in each grid cell varies.
In dccrg this is handled by each instance of the user's cell data class providing a MPI datatype corresponding to the data to be sent from or received by that particular cell.
An example of this is presented in Section \ref{sec:example_program}.

Completely arbitrary cell data can also be transferred between processes if the cell data class provides a serialize function which the MPI bindings of boost library will use for transferring cell data between processes\footnote{User-defined data types in http://www.boost.org/doc/libs/1\_49\_0/doc/html/mpi/tutorial.html}.
Although this method of transferring data between processes the most general it is also the slowest since data is first copied into a contiguous buffer by serialization and subsequently transferred by MPI resulting in at least one additional copy the data being created compared to pure a MPI transfer.
This is also the case in the SAMRAI framework \cite{wissink01} which supports transferring arbitrary patch data using the same technique.

\subsubsection{Automatic remote neighbor updates}

Dccrg can automatically transfer cell data between processes both for remote neighbor data updates and load balancing using simple function calls.
Furthermore whenever cell data is sent between processes either one MPI message per cell can be used or, similarly to \cite{wissink01}, all cells being sent to another process can be transferred using a single MPI message.
Updating the remote neighbor data between processes is possible using several methods.
The simplest one is the synchronous update function that updates the remote neighbor data between processes and returns once transfers have completed (see Section \ref{sec:example_program}).
The most fine-grained communication currently supported can be used by calling a separate function for initiating transfers and functions that wait for the sends and receives to complete respectively.
A typical usage scenario would consist of the following:
\begin{enumerate}
\item Start transferring remote neighbor data
\item Solve the inner cells of the simulation (cells without remote neighbors)
\item Wait for the data from other processes to arrive
\item Solve the outer cells of the simulation (cells with at least one neighbor on another process)
\item Wait for the data from this process to be sent
\end{enumerate}
The MHD scalability tests we present in section \ref{sec:mhd_tests} use this procedure with the exception that step 5 is executed before step 4 due to the technical implementation of the GUMICS-4 MHD solver.

\subsubsection{User-selectable neighborhood size}

As mentioned in Section \ref{sec:introduction} the size of cells' neighborhood in simulations is highly problem / solver dependent.
Specifically the problem / solver used in the simulation dictates the distance from which data is required at a cell in order to advance the simulation for one time step or one iteration in that cell.
In many previously published grid codes the size of cells' neighborhood is restricted to 1 either explicitly or implicitly.
For example in \cite{keppens12} in three dimensions a block (a single cell from the point of view of the grid) has 6 neighbors and it is assumed that a block consist of such a number of simulation cells that, for example, a solver needing data from a distance of 3 simulation cells can obtain that data from the neighboring block.
Other examples are \cite{burstedde11}, \cite{hariharan05}, \cite{khokhlov98}, \cite{kirk06}, \cite{macneice00} and \cite{stout97}.
Dccrg supports an arbitrarily large neighborhood chosen by the user when the grid is initialized.
The size of the neighborhood can be any unsigned integer and all other cells within that distance of a cell (in units of size of the cell itself) will be considered as a neighbors of the cell.
This enables the use of high-order solvers with the added possibility of refining each neighboring cell individually.
Naturally one can also store a sufficiently large block of simulation cells in one dccrg cell allowing one to use a small 6 cell neighborhood as done in \cite{keppens12}.
Zero neighborhood size is a special case in dccrg signifying that only face neighbors of equal size are considered neighbors (with AMR all of the refined neighbor's children are considered instead).
For example in a periodic grid without AMR neighborhood sizes of 1 and 2 would result in 26 and 124 neighbors per cell respectively.

Naturally the size of the neighborhood affects the amount of data that must be transferred between processes during remote neighbor data updates regardless of the grid implementation thus affecting parallel scalability.
Additionally since in dccrg a maximum refinement level difference of one is enforced between neighboring cells the size of the neighborhood does affect for example the amount of induced cell refinement (see Section \ref{sec:amr}).

\subsubsection{Ease of use}
\label{sec:example_program}

\begin{figure}[t]
\vspace*{2mm}
\begin{center}
\includegraphics[width=\columnwidth]{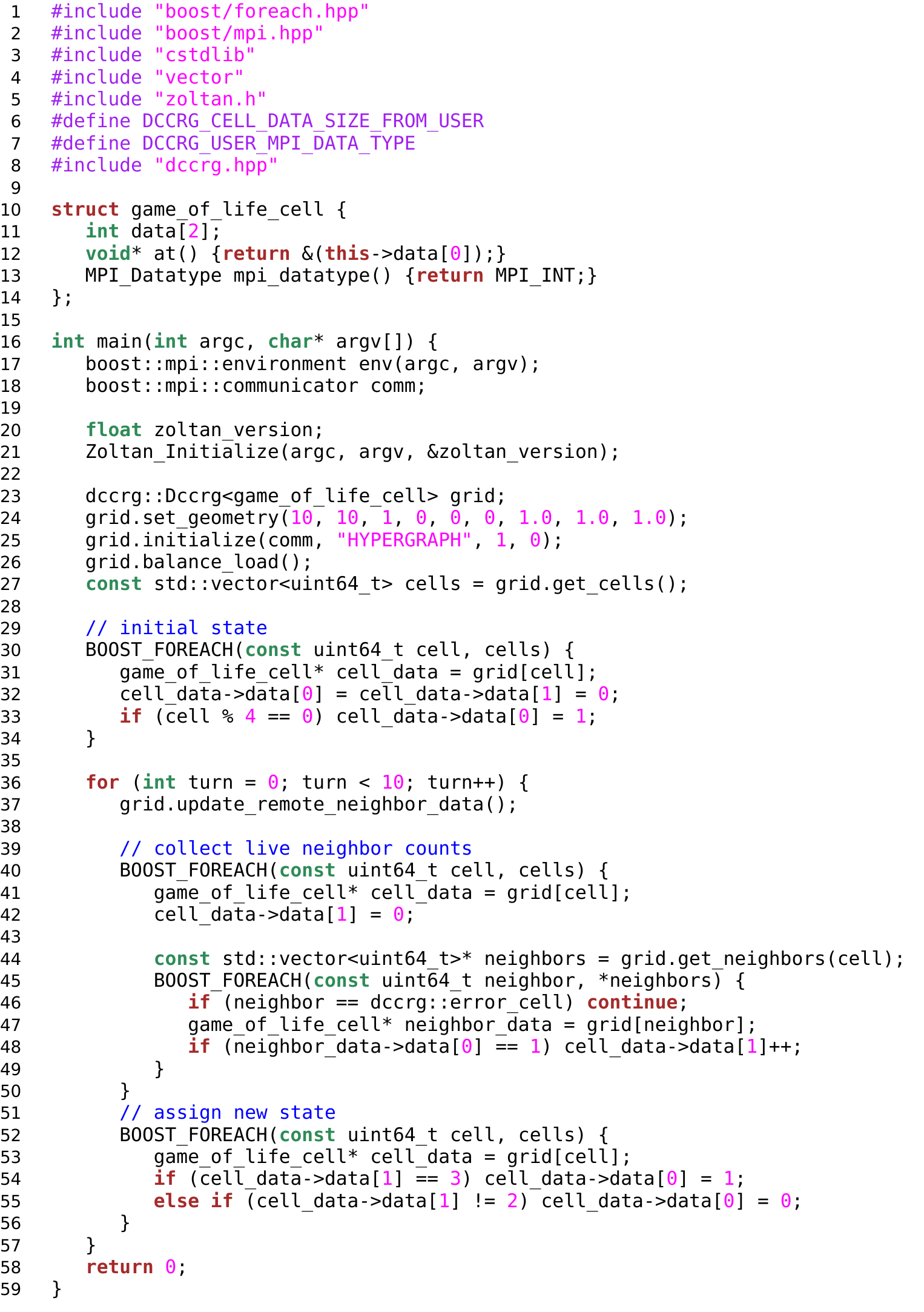}
\end{center}
\caption{
A complete parallel program playing Conway's Game of Life using dccrg, see the text for details.
}
\label{fig:gol}
\end{figure}

Even though initially dccrg was developed only for in-house use, it was nevertheless designed to be simple to use for the kinds of simulations it is targeted for.
Figure \ref{fig:gol} shows an example of a complete parallel program playing Conway's Game of Life using dccrg written in less that 60 lines of code (LOC) including whitespace and comments.
Lines 10...14 of the program define the class to be stored in every cell of dccrg along with member functions {\tt at} and {\tt mpi\_datatype} which dccrg calls when querying the information required for transferring cell data between processes.
The current state of a cell is saved into {\tt data[0]} and the number of its live neighbors is saved into {\tt data[1]}.
On line 23 an instance of the grid is created with the class defined above as cell data.
On line 24 the geometry of the grid is set to 10x10x1 cells at refinement level 0 with minimum coordinate at (0, 0, 0) and cells of size 1 in each dimension.
On line 25 the grid is initialized by setting the load balancing function to use in Zoltan, the neighborhood size and the maximum refinement level of cells.
Lines 26 and 27 balance the load using Zoltan and collect the local cells.
In this example the load is balanced only once and the local cell list does not change afterwards.
On line 46 the non-existing neighbors of local cells are skipped.
This is because the grid is initialized with non-periodic boundaries and neighbors that would be outside of the grid do not exist.

\begin{figure}[t]
\vspace*{2mm}
\begin{center}
\includegraphics[width=\columnwidth]{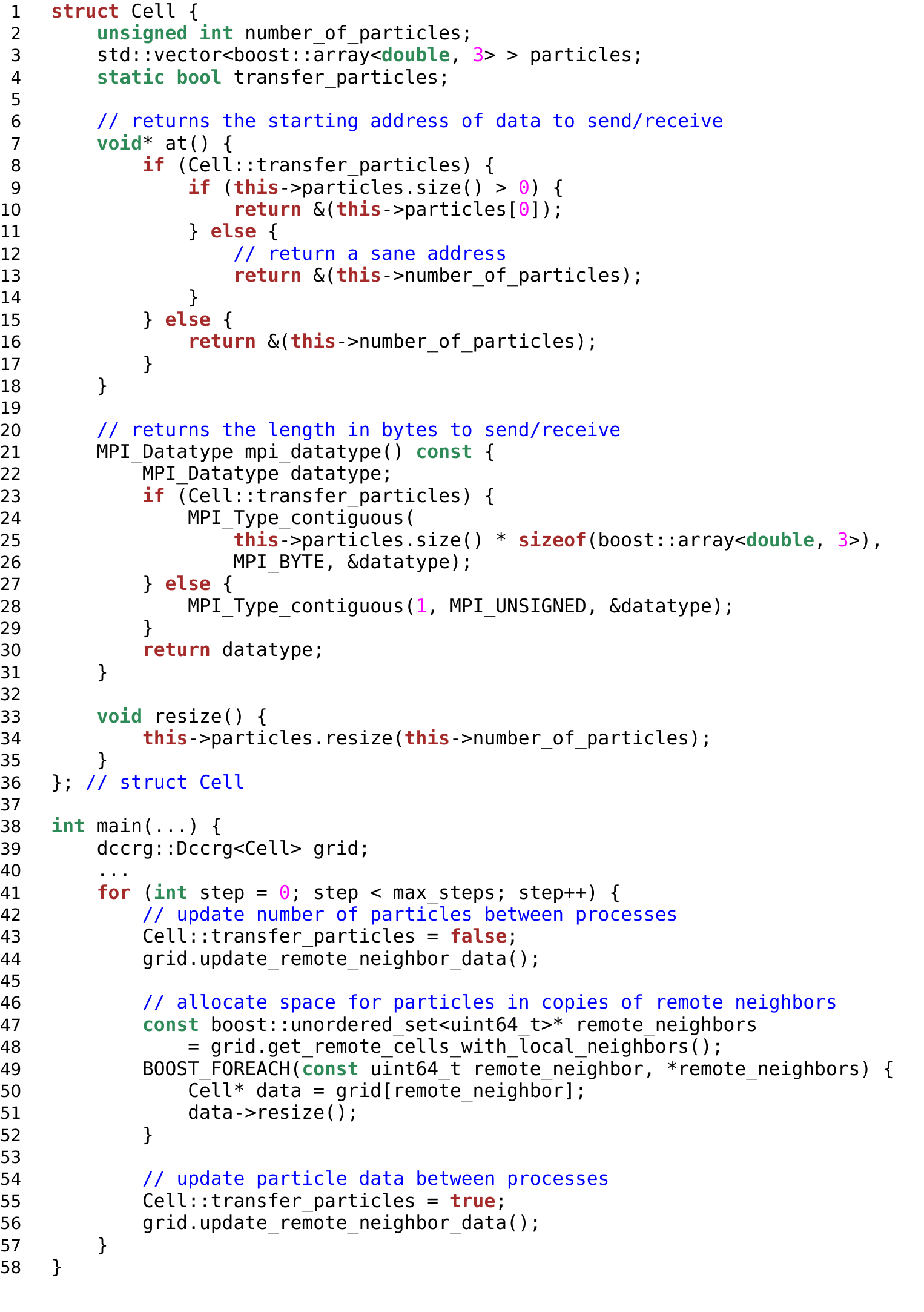}
\end{center}
\caption{
Relevant excerpts from a simple kinetic simulation showing how to use dccrg when the amount of data in grid cells varies both in space and in time, see text for details.
}
\label{fig:particles}
\end{figure}

Figure \ref{fig:particles} shows relevant excerpts from a simple kinetic simulation showing the use of dccrg in the case of variable amount of cell data, the full program can be viewed in the dccrg git repository\footnote{https://gitorious.org/dccrg/dccrg/blobs/master/tests/particles/simple.cpp}.
The remote neighbor update logic in the main simulation loop consists of the following steps:
\begin{enumerate}
\item The total number of particles in each cell is transferred between processes (lines 43 and 44)
\item Space for receiving particle data is allocated in local copies of remote cells based on their received total number of particles in step 1 (lines 47...52)
\item The particle coordinates are transferred between processes (lines 55 and 56)
\end{enumerate}
The cell data class of the example kinetic simulation must provide the correct information to dccrg when updating remote neighbor data:
The {\tt at} and {\tt mpi\_datatype} functions now return a different address and number of bytes respectively depending on whether the total number of particles or the particle coordinates are transferred between processes.
This is decided by the user in the main simulation loop.
Additionally the {\tt resize} function of the cell data class allocates memory for as many particles as is there are in {\tt number\_of\_particles}.
A similar approach to the one described above is also used in our Vlasov simulation (further developed from \cite{minna12}) where each real space cell has a separate adaptable velocity grid for ions consisting of a variable number of $4^3$ cell velocity blocks.

In the previous example two communications are required per time step because processes receiving particle data do not know the number of incoming particles in advance.
Since the MPI standard requires that the maximum amount of data to be received is known before calling the receive function the number of particles has to be communicated separately.
This guarantees that processes receiving particles can specify the size of the data to MPI and allocate the memory required for received particles.

\subsection{Load balancing / cell partitioning}

Load balancing is also accomplished easily with dccrg.
A user can call the {\tt balance\_load} function to let the Zoltan \cite{devine02} library create a new partition, and single cells can also be moved manually between processes using the {\tt pin} and {\tt unpin} functions.
Most of Zoltan's load balancing methods\footnote{http://www.cs.sandia.gov/Zoltan/ug\_html/ug\_alg.html\#LB\_METHOD} can be used, namely: NONE, RANDOM, BLOCK, RCB, RIB, HSFC, GRAPH, HYPERGRAPH and HIER.
In any case dccrg will transparently execute the new partition by transferring the necessary cell data between processes using MPI.

The structure of the grid in dccrg includes the owner of a cell in addition to the unique id of the cell (id is the key and owner is the value in a hash table).
Thus the cell ids themselves are not used for partitioning cells between processes and any cell can be moved to any process (for example by using the RANDOM partitioner of Zoltan which we have found to be very useful for testing).

\subsection{Adaptive mesh refinement algorithm}
\label{sec:amr}

Due to the unique mapping of cells' ids and their physical location and size it is straightforward to refine any given cell in the grid, i.e. to calculate the ids of the children of any cell, and can be done locally (see Section \ref{sec:hash_table}).
In order to enforce a maximum refinement level difference of one between neighboring cells whenever a cell is refined the refinement level of all neighbors is checked.
If the refinement level of any neighbor is less than that of the cell being refined that neighbor is also refined.
This is continued recursively until no more cells need to be refined.
The size of the cells' neighborhood affects induced refinement indirectly by changing the number of neighbors a cell has and hence the potential number of cells whose refinement will be induced.
In dccrg a few simplifications have been made to AMR:
1) Any set of cells can only be refined once before calculating induced refinement (by {\tt stop\_refining}), e.g. induced refinement can only increase the refinement level of cells by one, and
2) Unrefining a cell does not induce unrefinement, e.g. any cell which has at least one neighbor with refinement level larger than the cell being unrefined (i.e. it has a smaller neighbor) cannot be unrefined.

In a parallel setting the only difference to the above is that whenever a process refines or unrefines a cell that information has to be given to all processes which have cells within a certain distance of the cell that was refined or unrefined.
Currently this distance is equal to infinity, e.g. all changes to the structure of the grid (refines and unrefines) are communicated globally.
This has a significant impact on the parallel scalability of AMR in dccrg and is discussed in Section \ref{sec:tests} and a method for making this minimum distance finite is discussed in Section \ref{sec:discussion}.
The changes in grid structure are exchanged between processes after each recursive step of induced refinement which are continued until no more cells need to be refined.

\subsection{Parallel implementation}

Good scalability in distributed memory machines requires asynchronous point-to-point MPI communication between processes with minimal total amount of communication, and especially minimal amount of global communication.
The number of MPI messages should also be minimized in order not to burden the network with unnecessary traffic.
Therefore any process must know which processes require data from local cells and from which remote cells data is required during remote neighbor updates without querying that information from other processes beforehand.
Thus in dccrg every process knows the structure of the grid (e.g.~which cells exist and which processes own them) to at least a certain distance from any of its cells so it can calculate locally which cells' data to send and receive from other processes.
Due to this dccrg does not require global communication during ordinary time stepping, e.g.~remote neighbor data updates, which enables excellent scalability when not doing load balancing or AMR as shown in section \ref{sec:mhd_tests}.
Internally dccrg precalculates these send and receive lists whenever the structure of the grid changes due to AMR or cells being moved between processes.

Even though the replicated mesh metadata of dccrg does not include the data stored in each cell it nevertheless limits the size of dccrg grids in practice to less than 100 M existing cells.
This number does not depend on the refinement level or physical location of the cells and has so far been more than sufficient for our needs.
To our knowledge the only parallel grid library that does not have any persistent global data structures is \cite{sundar08}.
In \cite{burstedde11} and \cite{bangerth11} only the macrostructure of the grid (i.e. cells of refinement level 0) is known by all processes.
According to the authors this limits the size of the grid to the order of $10^5$\ldots$10^6$ cells of refinement level 0 but does not limit the number of smaller cells.

\section{Scalability results}
\label{sec:tests}

The time stepping scalability of dccrg (e.g.~without AMR or load balancing) depends mostly on the hardware running the simulation and on three parameters specific to each simulated system:
1) the time required to solve the inner cells of a process,
2) the amount of data transferred during the remote neighbor data update of a process and
3) the time required to solve the outer cells of a process.
We show the dependency of dccrg scalability on these parameters by varying the total number of cells and processes and by using a MHD solver in one, two and three dimensions.
The run-time AMR scalability of dccrg is also presented.

The non-AMR scalability tests were carried out on three different supercomputers:
1) A 2 k core Cray XT5m system with 12 core nodes connected by SeaStar2 installed at the Finnish Meteorological Institute which we will refer to as Meteo,
2) A 295 k core IBM Blue Gene/P system with 128 core nodes (nodeboards) installed at the J\"ulich Supercomputing Centre which we will refer to as Jugene, and
3) A 12 k core bullx system with 32 core nodes connected by InfiniBand installed at the Tr\`es Grand Centre de Calcul which we will refer to as Curie.

\subsection{MHD tests without AMR}
\label{sec:mhd_tests}

First we show the time stepping scalability of dccrg in several MHD problems with a solver developed for the global MHD model GUMICS-4 (\cite{pekka12}) which solves ideal MHD equations in conservative form.
Specifically the solver is a first order Roe's approximate Riemann solver for a Godunov type problem \cite{roe81}.
In the test results we present here only the Roe solver from GUMICS-4 is used as we do not experience problems with negative pressures or densities in the presented tests.
The nature of the solver is such that when solving the flux through a face data is only required from cells adjacent to the face, irrespective of the size of cells involved.
Thus interpolation of data is not required at any point in the solution and if a cell has more than one face neighbor in any direction the flux through each common face is solved in the usual way.
In the tests presented here we do not include a background magnetic field which is used in GUMICS-4 to represent the Earth's dipole field.

The problems we use are the one-dimensional shock tube presented for example in \citep{torrilhon02}, the two-dimensional circularly polarized Alfv\'en wave presented by \citep{toth00} and further elaborated on by \citep{gardiner05}, and the three-dimensional blast wave presented by \citep{gardiner08}.
Periodic boundary conditions are used in all tests, except for the shock tube test in the direction of the tube where initial conditions are enforced after every time step.
In MHD tests every cell contains the cell-averaged values of the conservative MHD variables (density, momentum density, total energy density and magnetic field) giving a total of 128 bytes which must be transferred when updating the data of one cell between two processes.
Since only the face neighbors of a cell are required for calculating the next time step we use a neighborhood size of zero in dccrg.
In these tests processes execute one collective MPI communication per time step in order to dynamically calculate the maximum physical length of the time step.
No other global communication is done.
Since the grid is static in these tests the computational load is balanced only once at the start of the simulation by using a Hilbert space-filling curve\footnote{https://gitorious.org/sfc/sfc/blobs/master/sfc++.hpp} instead of Zoltan.

\begin{figure}[t]
\vspace*{2mm}
\includegraphics[width=17cm]{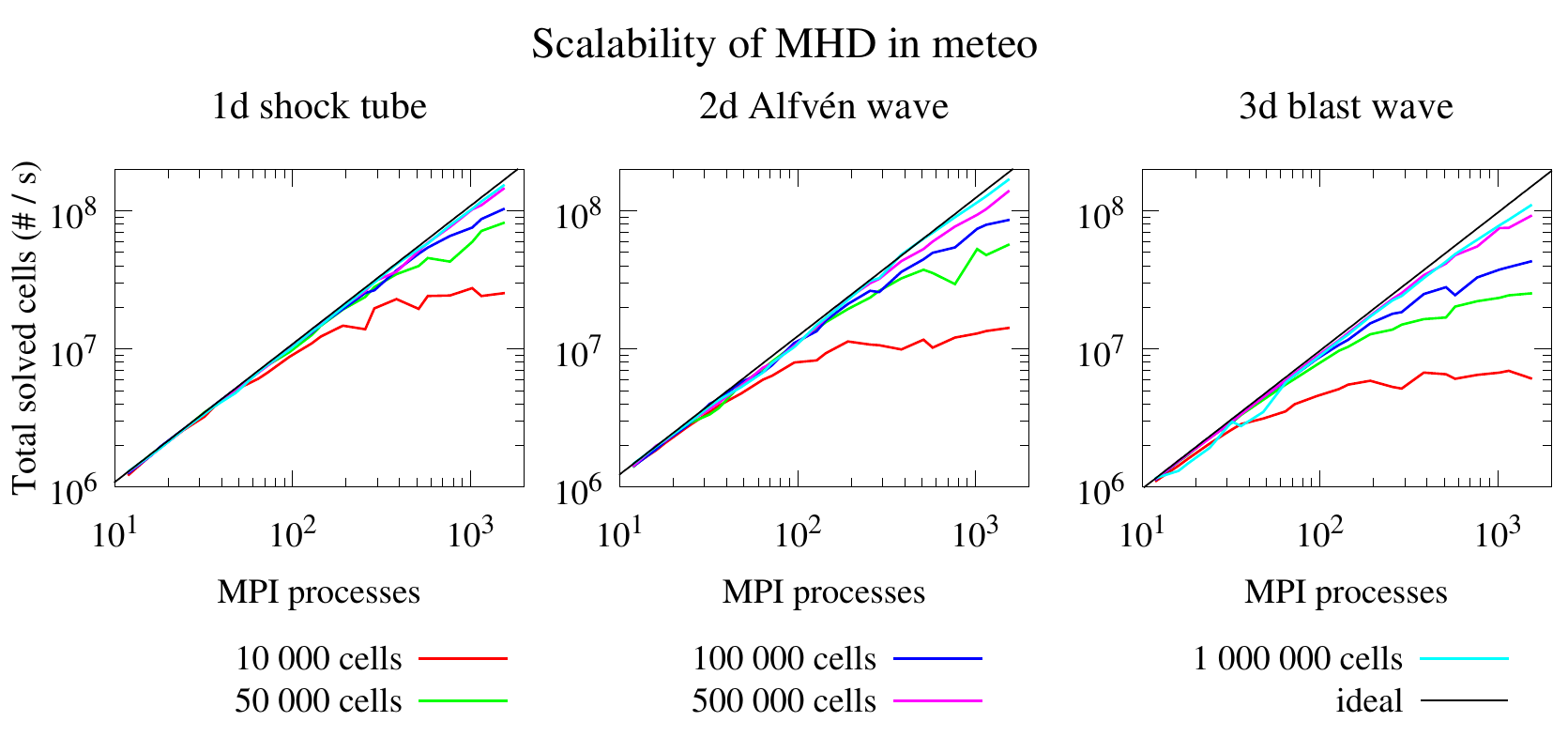}
\caption{
Strong MHD scalability tests results with a static grid in one, two and three dimensions in Meteo (Cray XT5m).
Total number of solved cells per second is shown as a function of the number of processes used and the total number of cells in each simulation.
Ideal line is extrapolated from the 12 process result with 1 M cells.
}
\label{fig:mhd-meteo}
\end{figure}

Figure \ref{fig:mhd-meteo} shows the results of strong scalability tests using MHD with a static grid in Meteo in one, two and three dimensions.
The total number of cells (10 k, 50 k, 0.1 M, 0.5 M and 1 M) is kept constant while the number of MPI processes is increased from 12 to 1536.
In each test case scalability improves with the total number of cells because processes have more inner cells to solve while remote neighbor data is being transferred.
For example in the shock tube test every process requires the data of two remote neighbors at most while the number of inner cells with 1.5 k processes increases from about 4 (10 k total cells) to 649 (1 M total cells).
With 1 M cells the one and two dimensional tests scale almost ideally in Meteo and the three dimensional test is also quite close to ideal.
The overall decrease in scalability with increasing number of dimensions is due to more data being transferred between processes for the same number of local cells.

\begin{figure}[t]
\vspace*{2mm}
\includegraphics[width=17cm]{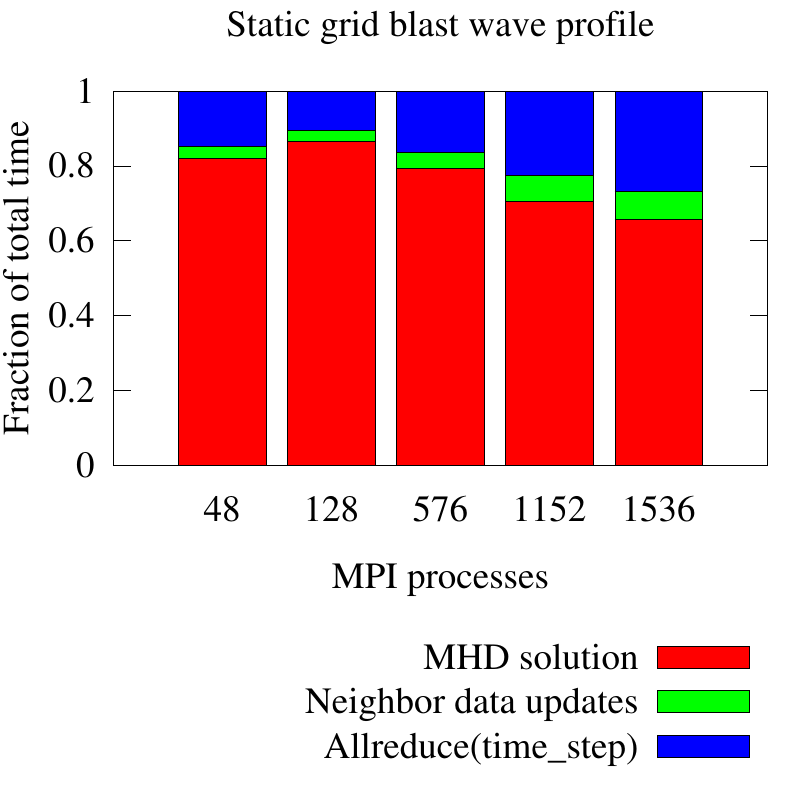}
\caption{
Profile of the three dimensional MHD blast wave test without AMR using 1 M cells.
Allreduce labels the only global communication executed each time step where the maximum length of the physical time step is obtained using {\tt MPI\_Allreduce}.
Initialization and file I/O are not included.
}
\label{fig:meteo-profile}
\end{figure}

As suggested by the scalability results above most of the simulation time is spent solving MHD which is shown in Figure \ref{fig:meteo-profile} for the three dimensional blast wave test using 1 M cells.
The only global communication executed per time step while simulating is the calculation of the maximum length of the physical time step and is obtained using {\tt MPI\_Allreduce}.
This is labeled as Allreduce in Figure \ref{fig:meteo-profile} and basically shows the computational and MPI imbalance between processes due to load balancing.
Initialization and file I/O are not included in the profile and other parts of the simulation take an insignificant fraction of the total run time.

\begin{figure}[t]
\vspace*{2mm}
\begin{center}
\includegraphics[width=\columnwidth]{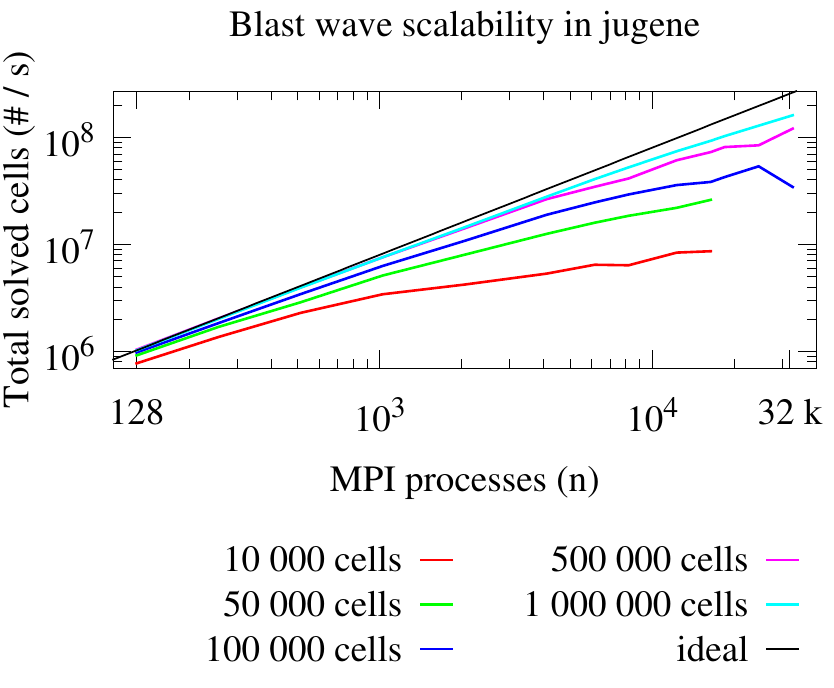}
\end{center}
\caption{
Strong MHD scalability tests results with a static grid in three dimensions in Jugene (IBM Blue Gene/P).
Total number of solved cells per second is shown as a function of the number of processes used and the total number of cells in each simulation.
Ideal line is extrapolated from the 128 process result with 1 M cells.
}
\label{fig:blast-wave-jugene}
\end{figure}

\begin{figure}[t]
\vspace*{2mm}
\begin{center}
\includegraphics[width=\columnwidth]{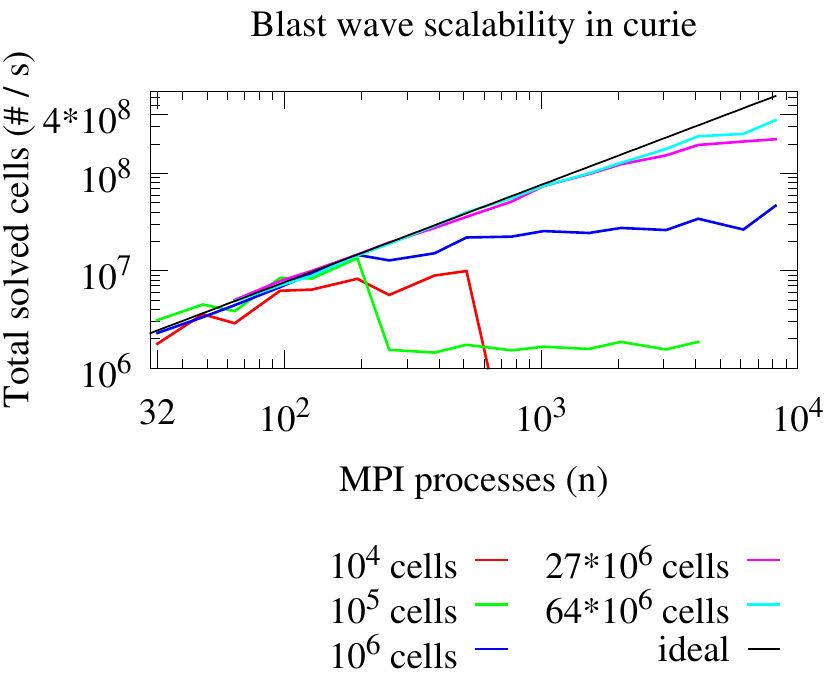}
\end{center}
\caption{
Strong MHD scalability tests results with a static grid in three dimensions in Curie (bullx InfiniBand).
Total number of solved cells per second is shown as a function of the number of processes used and the total number of cells in each simulation.
Ideal line is extrapolated from the 32 process result with 64 M cells.
}
\label{fig:blast-wave-curie}
\end{figure}

The non-AMR scalability tests were also carried out in Jugene and Curie and the results for the three-dimensional blast wave are shown in Figures \ref{fig:blast-wave-jugene} and \ref{fig:blast-wave-curie} respectively.
Similarly to Meteo the one and two dimensional tests (not shown) scale better than the three-dimensional test in both Jugene and Curie.
The overall results are similar in all tested machines, e.g.~scalability improves with increasing number of total cells and decreasing number of dimensions.
In Jugene very good scalability up to 32 k processes is obtained for a total number of cells of 1 M and above.
The total simulation speed in Jugene is only slightly above that of Meteo mostly due to the relatively small single-core performance of Jugene.
Additionally the average number of cells per process is more than 20 times larger in Meteo than in Jugene for the maximum number of processes used but this has only a small effect on scalability in Jugene.

In Curie good scalability up to 8 k processes is obtained only with 64 M total cells but with a maximum solution speed of nearly 400 M solved cells per second which is over twice of that in Jugene.
We attribute this to the relatively low node interconnect and high single-core and performance of Curie respectively when compared to Jugene.

\subsection{Scalability of run-time AMR}
\label{sec:amr_scalability}

\begin{figure}[t]
\vspace*{2mm}
\begin{center}
\includegraphics[width=\columnwidth]{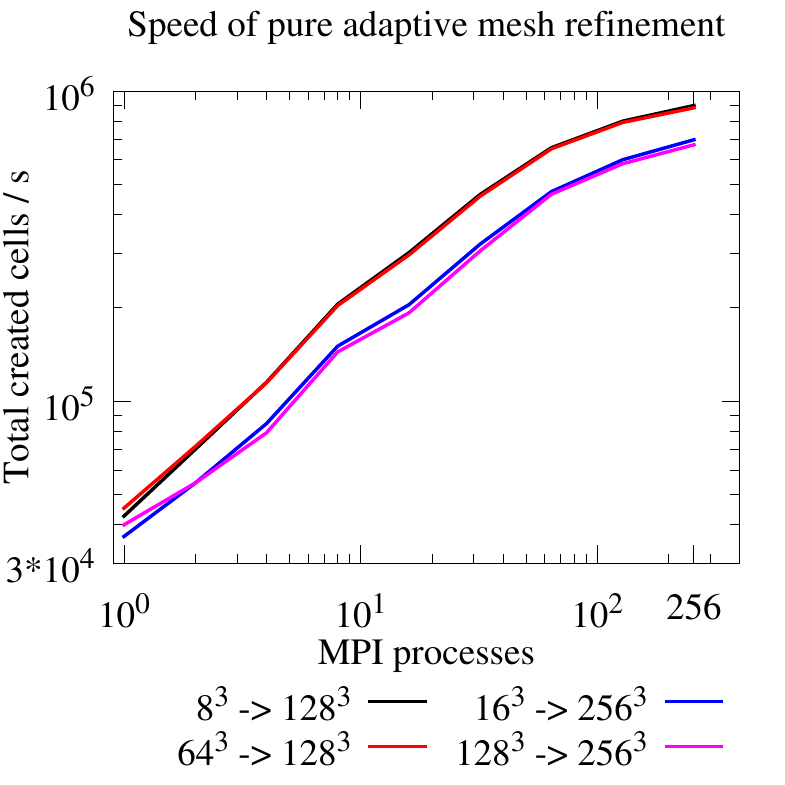}
\end{center}
\caption{
Speed of adaptive mesh refinement in dccrg.
Initial size of the grid is $8^3, 16^3, 64^3$ or $128^3$ cells.
Every process refines each local cell until the total size of the grid is $128^3$ or $256^3$ cells.
}
\label{fig:amr_speed}
\end{figure}

Figure \ref{fig:amr_speed} shows the speed of pure adaptive mesh refinement in dccrg.
Initially the grid is $8^3, 16^3, 64^3$ or $128^3$ cells and every process refines all local cells until the total size of the grid is $128^3$ or $256^3$ cells.
Initially the cells were partitioned using a space-filling curve and this is not included in the timings.
Cells also were not transferred between processes during AMR.
As can be seen in Figure \ref{fig:amr_speed} the maximum cell refining speed of dccrg is of the order of 1 M cells per second.
The linear scalability of AMR up to some 32 MPI processes is explained by the fact that after changes in the structure of the grid the arrow lists are recalculated only for local cells and MPI communication has not yet become a bottleneck.
At 256 processes the amount of global communication required for updating the structure of the grid between all processes starts to significantly affect the speed of AMR.
This is discussed further in Section \ref{sec:amr_blast_wave}.

\subsection{Scalability of blast wave test with AMR}
\label{sec:amr_blast_wave}

Here we present the scalability of dccrg with AMR in the three-dimensional blast wave test used in Section \ref{sec:mhd_tests}.
In this test a procedure similar to the one in GUMICS-4 (eq.~2 in \cite{pekka12}) is used to decide whether to refine or unrefine a cell:
A refinement index is calculated for each cell based on the relative difference of several variables between a cell and its face neighbors.
Here the calculation of refinement index $\alpha$ additionally includes velocity shear relative to the maximum wave velocity from the cells' interface.
The full equation for the refinement index $\alpha$ is:
$$
\alpha = \mathrm{max}(
\frac{\Delta \rho}{\widehat{\rho}},
\frac{\Delta U_1}{\widehat{U_1}},
\frac{(\Delta \boldmath{p})^2}{2 \widehat{\rho U_1}},
\frac{(\Delta \boldmath{B}_1)^2}{2 \mu_0 \widehat{U_1}},
\frac{|\Delta \boldmath{B}_1|}{\widehat{B_1}},
\frac{(\Delta \boldmath{v})^2}{v_{min}}
)
$$
where $\Delta$ denotes the difference in a variable between two cells, the hat denotes a minimum of the two values (as it actually does also in \cite{pekka12}), $v_{min} = \widehat{\boldmath{v}^2} + (0.01 \cdot v_{wave})^2$ and $v_{wave}$ is the maximum wave velocity from the cells' interface.
In this test a cell is refined if $\alpha > 0.02 \cdot (l+1) / L$, where $l$ is the cell's current refinement level and $L = 4$ is the maximum refinement level of the grid.
In other words a cell is refined to the maximum refinement level if its refinement index exceeds 0.02.
A cell is unrefined if $\alpha < 0.02 \cdot (l+1)/ L / 2$, e.g.~a cell is kept at refinement level 0 if $\alpha < 0.0025$ and none of the cell's neighbors' refinement levels exceed 1 (due to dccrg enforcing a maximum refinement level difference of one between neighbors).

\begin{figure}[t]
\vspace*{2mm}
\begin{center}
\includegraphics[width=\columnwidth]{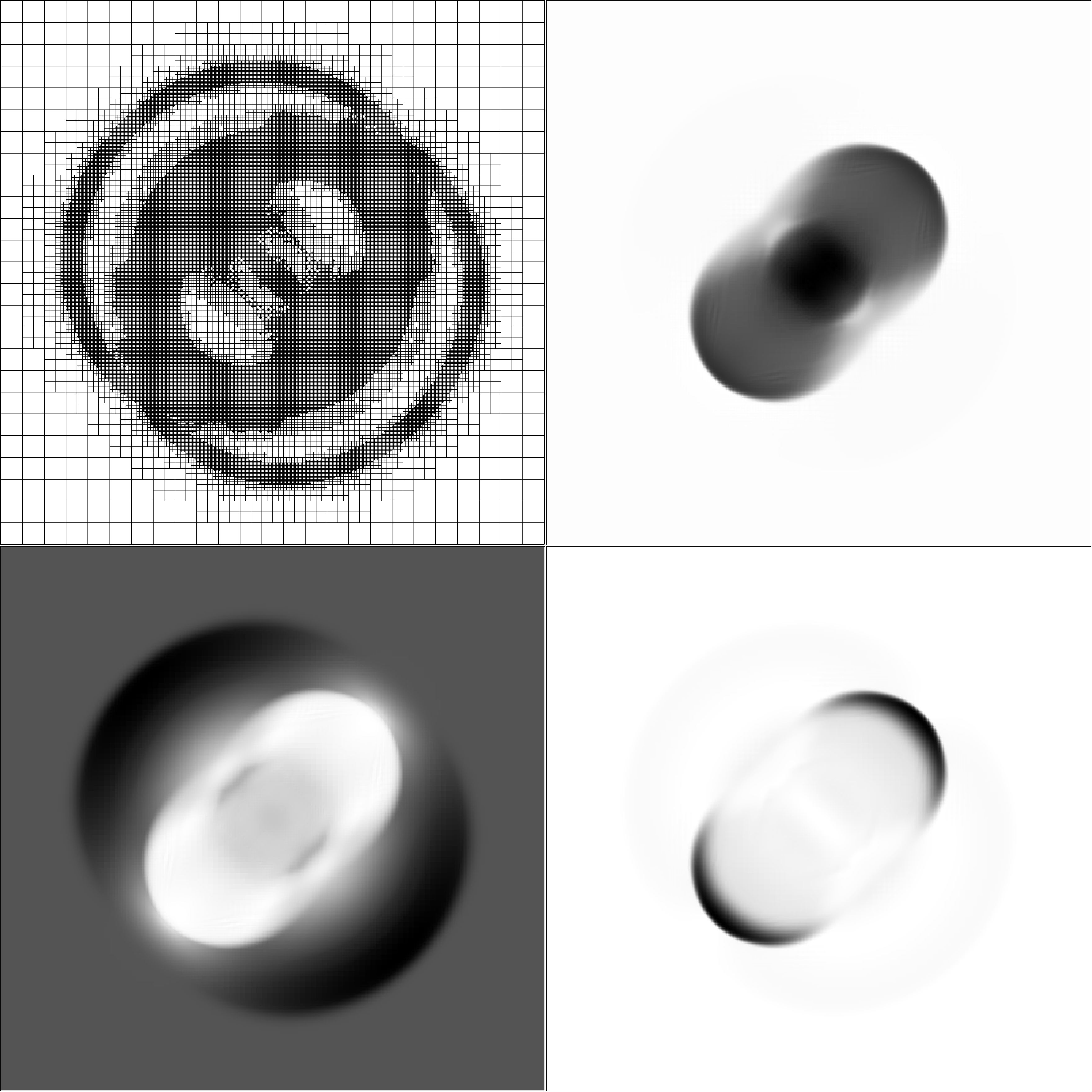}
\end{center}
\caption{
Adaptive mesh refinement used in a MHD blast wave test (from \cite{gardiner08}) showing from left to right, top to bottom the grid, pressure, magnetic and kinetic energy density during the simulation (final state of simulation in the print version) in the y = 0 plane when grid is adapted at every time step.
At the end of the simulation the fraction of maximum to minimum values are 15 for density (not shown), 43 for pressure and 2.3 for magnetic energy density.
}
\label{fig:amr_result}
\end{figure}

We use a maximum refinement level of 4 in this test with an initial grid of $25^3$ cells which results in an effective resolution of $400^3 = 64$ M cells.
The computational load is balanced using Zoltan's recursive coordinate bisection (RCB) algorithm whenever the fraction of local cells ($f_c = N_{max} / N_{min}$, where max and min are the maximum and minimum number of local cells among all processes respectively) exceeded a specified limit.
Animation 1 (Figure \ref{fig:amr_result} in the print version) shows from left to right, top to bottom the grid, pressure, magnetic and kinetic energy density during the simulation (at the end of the simulation in print version) in the y = 0 plane when grid is adapted at every time step.
At the end of the simulation the fraction of maximum to minimum values are 15 for density (not shown), 43 for pressure and 2.3 for magnetic energy density.
Even though the MHD solver we use is simpler than the one in \cite{gardiner08} the results are still close due to the high effective resolution achieved by using run-time AMR.

\begin{figure}[t]
\vspace*{2mm}
\begin{center}
\includegraphics[width=\columnwidth]{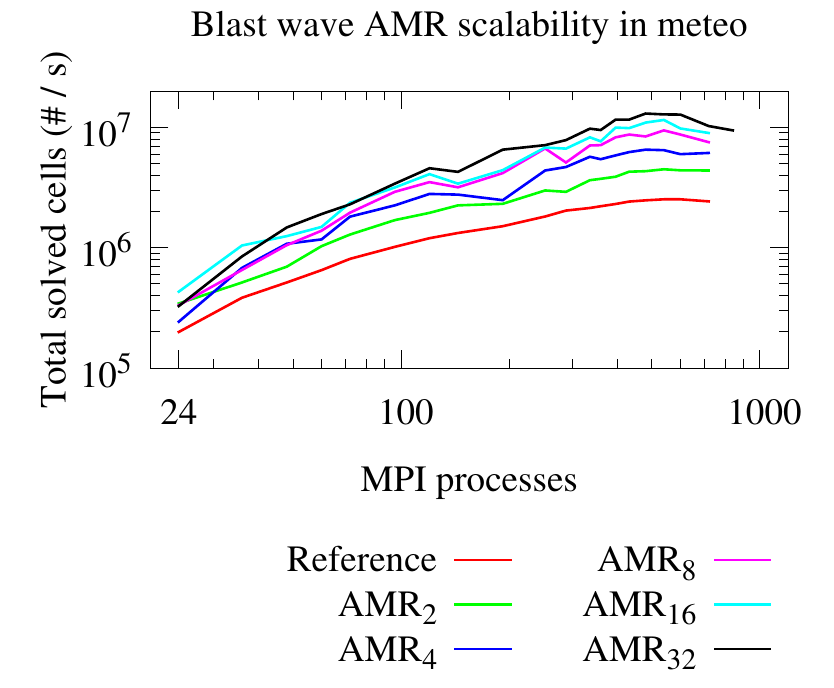}
\end{center}
\caption{
Scalability of adaptive mesh refinement used in a MHD blast wave test (\cite{gardiner08}).
In the reference run a CFL of 0.4 is used, AMR is done at every time step and the load is balanced whenever the local cell fraction $f_c$ exceeded 2, where $f_c = N_{max} / N_{min}$ and max and min are the maximum and minimum number of local cells among processes respectively.
The $\mathrm{AMR}_N$ runs are otherwise identical to the reference run but CFL is set to $0.4 / N$ and AMR is done every Nth time step, essentially multiplying the amount of non-AMR work in these runs by N.
}
\label{fig:amr_scalability}
\end{figure}

Figure \ref{fig:amr_scalability} shows the total solution speed during the simulations as a function of the number of MPI processes used.
In the reference run a CFL \cite{courant67} of 0.4 is used, AMR is done at every time step and the load is balanced whenever the local cell fraction $f_c \geq 2$.
The $\mathrm{AMR}_N$ runs are otherwise identical to the reference run but CFL is set to $0.4 / N$ and AMR is done every Nth time step, essentially multiplying the amount of non-AMR work in these runs by N.
The results between different AMR runs are identical by visual inspection except for increased diffusion in runs with low CFL.
The ratio of work required by AMR and the rest of the simulation has a significant effect on the total solution speed.
The solution speed is a factor of 5 higher in the $\mathrm{AMR}_{32}$ run than in the reference run when using about 500 MPI processes.
In the reference AMR run the total solution speed is close to $1/10$ of the non-AMR version with up to 144 processes and in the $\mathrm{AMR}_{32}$ the speed is close to $1/3$ with up to 288 processes after which both fractions start to decrease.
We define these as the regions of excellent AMR scalability.
On the other hand in all of the AMR runs the total solution speed increases up to about 500 to 600 processes after which it starts to decrease.
We define this as the region where AMR is scalable.
The total number of cells in the AMR runs averages to 4.5 M which is about $1/14$ of the non-AMR version.
Consequently in the region of excellent scalability the time to solution when using AMR is about 67 \% to 22 \% of the non-AMR time for the reference and $\mathrm{AMR}_{32}$ runs respectively.
Even with a higher number of MPI processes it can still be advantagous to use AMR because the number of simulation cells is over a magnitude lower than without AMR.
At the end of the AMR runs 9.9 M cells exist in the grid and the total number of cells created and removed is between 40.2 M and 40.7 M, depending mostly on the diffusion, and averages to about 91 k added + removed cells per time step.

\begin{figure}[t]
\vspace*{2mm}
\includegraphics[width=17cm]{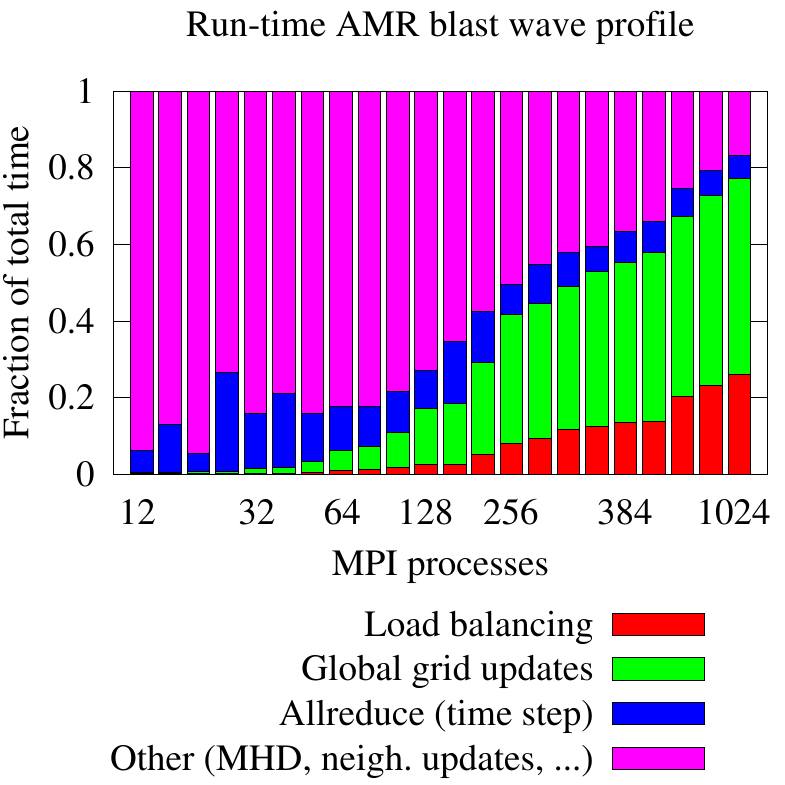}
\caption{
Profile of the three dimensional MHD blast wave test with AMR using at most 10 M cells.
Allreduce labels the calculation of the maximum length of the physical time step obtained using {\tt MPI\_Allreduce}.
Initialization and file I/O are not included.
}
\label{fig:meteo-profile-amr}
\end{figure}

Figure \ref{fig:meteo-profile-amr} shows which parts of the AMR blast wave test require the most time.
As the number of processes is increased the largest fraction of simulation time is spend in global communication related to AMR and load balancing.
The Allreduce label again indicates global calculation of the physical time step and the Load balancing label indicates the simulation time spent in load balancing related functions.
At about 300 MPI processes and above the largest fraction of simulation time is spent communicating changes in the structure of the grid between all processes.
This includes both refining and unrefining cells as well as load balancing and in each case the {\tt MPI\_Allgatherv} function is used for distributing the changes in grid structure between all processes.
Only a small fraction of the time spent in load balancing related functions is taken by Zoltan.

\section{Discussion}
\label{sec:discussion}

While the ease of use of a software library is subjective it can be quantified by the number of lines of code required for usage and compared against other libraries when using the same programming language.
With dccrg a complete parallel program playing Conway's game of life can be implemented in less than 60 LOC including whitespace and comments.
Even though the required LOC is a crude estimate for a software library's ease of use it is nevertheless telling that such a short parallel program does not seem to be possible with other grid libraries.

The flexibility of dccrg also stands out since as far as we know only \cite{wissink01} allows one to easily exchange arbitrary cell data between MPI processes.
Additionally dccrg supports transferring user-defined MPI datatypes which is critical for performance in some applications.
For example when solving the 6 dimensional Vlasov equation in the Earth's magnetosphere (\cite{minna12}) the simulation is heavily memory bound and using MPI datatypes directly for exchanging remote neighbor data is significantly faster than serializing said data into an additional buffer(s) before transfer.
Dccrg also provides automatic neighbor data updates between processes with the ability of easily overlapping computation with communication.

Currently the largest drawback of dccrg is the fact that the entire structure of the grid is known by every process, i.e. a part of the mesh metadata is replicated by all processes.
The global data structure prevents grids larger than about 100 M cells but this has not been a problem for us and can be worked around by storing blocks instead of single cells into dccrg (similarly to \cite{holst07}, for example).
The global grid data structure of dccrg also reduces the scalability of AMR in the worst case to about 200 and overall to about 600 MPI processes.
Nevertheless using AMR can lead to significant savings in the required memory as the number of cells can be one or even two orders of magnitude lower.
Depending on the problem the required CPU time can also be significantly reduced when using AMR especially when the number of MPI processes used is of the order of 300 or less.
It should also be noted that using threads to parallelize solvers within a shared memory node could effectively multiply the scalability range of simulations by the number of cores within one node, but this was not investigated.

Removing or significantly reducing the global data structure (as done in \cite{sundar08}, \cite{burstedde11} and \cite{bangerth11}) should improve both the largest attainable grid size and scalability of AMR considerably.
Intuitively this is straightforward since with the exception of load balancing every process only needs to know the structure of the grid up to some finite distance from local cells.
In order to be able to arbitrarily refine and unrefine grid cells without global communication local changes in the structure of the grid must be communicated between all neighboring processes.
A neighboring process is defined as any process that has one or more of its cells inside the neighborhood of any cell of refinement level 0 that overlaps a local cell.
In other words if only level 0 cells exist in the grid then the owners of all remote neighbors of local cells are considered as neighboring processes;
and this holds no matter how the grid is subsequently refined and unrefined assuming that cells are not transferred between processes (load balancing).
Global communication can also be avoided during load balancing if, for example, cells can be transferred only between neighboring processes.
Even in this case new neighbor processes have to be recalculated but global communication is not required because cells could only have been transferred to/from a subset of all processes.
Implementing this completely distributed mesh metadata is left to a subsequent study.

We presented the distributed cartesian cell-refinable grid (dccrg): an easy to use parallel structured grid library supporting adaptive mesh refinement and arbitrary C++ classe as cell data.
Various MHD scalability results were presented and depending on the problem, hardware and whether AMR is used excellent to average scalability is achieved.
Dccrg is freely available for anyone to use, study and modify under version 3 of the GNU Lesser General Public License and can be downloaded from https://gitorious.org/dccrg.

\section{Acknowledgements}
This work is a part of the project 200141-QuESpace, funded by the European Research Council under the European Community's seventh framework programme.
The research leading to these results has also received funding under grant agreement no 260330 of the European Community's seventh framework programme.
IH and MP are supported by project 218165 and AS is supported by project 251797 of the Academy of Finland.
Results in this paper have in part been achieved using the PRACE Research Infrastructure resource Curie based in France at TGCC and Jugene based in Germany at JSC.
IH thanks Daldorff, L.K.S. and Pomoell, J. for insightful discussions.

\end{document}